\documentclass[preprint]{aastex_mst}
\usepackage{natbib}
\usepackage{mathrsfs}
\usepackage{rotating}
\usepackage{graphicx}
\usepackage{epstopdf}
\usepackage{lscape} 

\newcommand{\fn}[1]{\footnote{\scriptsize{#1}}}

\newcommand{\Eqn}[1]{Eq{#1}.}  
\newcommand{\Fig}[1]{Fig{#1}.}  

\shorttitle{Chaotic Diffusion of Resonant KBOs}
\shortauthors{Tiscareno and Malhotra}
\slugcomment{Accepted to AJ}

\begin{document}

\title{Chaotic Diffusion of Resonant Kuiper Belt Objects}
\author{Matthew S. Tiscareno$^1$ and Renu Malhotra$^2$}
\affil{$^1$ Department of Astronomy, Cornell University, Ithaca, NY 14853\\$^2$ Lunar and Planetary Laboratory, University of Arizona, Tucson, AZ 85721}
\email{matthewt@astro.cornell.edu}

\begin{abstract}

We carried out extensive numerical orbit integrations to probe the long-term chaotic dynamics of the two strongest mean motion resonances of Neptune in the Kuiper belt, the 3:2 (Plutinos) and 2:1 (Twotinos).  Our primary results include a computation of the relative volumes of phase space characterized by large- and small-resonance libration amplitudes, and maps of resonance stability measured by mean chaotic diffusion rate.  We find that Neptune's 2:1 resonance has weaker overall long-term stability than the 3:2---only $\sim 15\%$ of Twotinos are projected to survive for 4~Gyr, compared to $\sim 27\%$ of Plutinos, based on an extrapolation from our 1-Gyr integrations.  We find that Pluto has only a modest effect, causing a $\sim 4\%$ decrease in the Plutino population that survives to 4~Gyr. Given current observational estimates, and assuming an initial distribution of particles proportional to the local phase space volume in the resonance, we conclude that the primordial populations of Plutinos and Twotinos formerly made up more than half the population of the classical and resonant Kuiper Belt.  We also conclude that Twotinos were originally nearly as numerous as Plutinos; this is consistent with predictions from early models of smooth giant planet migration and resonance sweeping of the Kuiper Belt, and provides a useful constraint for more detailed models.
\end{abstract}


\section{Introduction}

Over the past 15 years of observations, it has become clear that the trans-Neptune population of Kuiper Belt objects (KBOs) consists of several distinct dynamical classes \citep{JL00,MDL00,GMV08}. The ``classical'' KBOs, found primarily in the semimajor axis range of 40~AU to 48~AU, consist of two sub-classes: a dynamically cold population having relatively circular and low-inclination orbits, and a dynamically hot population with higher orbital eccentricities and inclinations \citep{Brown01,LS01,TB02}.  The scattered disk objects (SDOs) have highly eccentric and inclined orbits, with perihelia a few~AU beyond Neptune's orbit \citep{DL97}; a small number of the SDOs with perihelia $>40$~AU are thought to be a distinct population, the ``extended scattered disk'' (also known as the `detached objects') \citep{Gladman02}.  Finally, there are the resonant KBOs, which have orbits in mean motion resonance with Neptune; approximately 23\% of all known KBOs are in this class \citep{Chiang07}. Most of the known resonant KBOs are in the 3:2 resonance at semimajor axis $a=39.4$~AU, and have been dubbed `Plutinos' in recognition of the largest and longest-observed member of this population; the 2:1 resonance at $a=47.7$~AU has the next-highest observed resonant population, and have been dubbed `Twotinos'; several other resonances are also populated \citep{Chiang03,Chiang07}.

The particular interest of this paper is the dynamics of bodies within the 3:2 and 2:1 mean-motion resonances with Neptune.  Because the structure of these resonances provides a readily understandable mechanism for preserving a population for 4~Gyr or more, as well as dynamical pathways to less stable orbits where they may encounter Neptune and be transported elsewhere in the solar system, resonant KBOs have been proposed as the source populations for the Centaurs and Jupiter-family comets \citep{DLB95,DL97,Morby97}. Resonant KBOs are also of great interest because their relative populations hold clues to the orbital migration history of the giant planets and the dynamical history of the outer solar system as a whole \citep{Malhotra95,MB04,MLG08}.

In the present work, we use numerical analysis to obtain a detailed map of stability as it varies with location in the phase space of the 3:2 and 2:1 mean-motion resonances with Neptune. One major goal of this study is to estimate (by extrapolating the rates seen in our 1-Gyr numerical integrations) the Plutino and Twotino populations $\sim4$~Gyr ago when the dynamical structure of the Kuiper belt was presumably established. We also investigate the effects of Pluto on the Plutino population, and the behavior of particles after they have escaped from resonance.  

Section~\ref{plutinos.Resonances} of this paper discusses the properties of resonant particles and our treatment of chaotic diffusion; Section~\ref{NumModel} describes the numerical experiments that we carried out to study the long-term stability of particles in the 3:2 and 2:1 resonances; Section~\ref{plutinos.Results} describes the results of our numerical models; Section~\ref{Discussion} provides discussion, analysis, and comparison to previous work; and Section~\ref{plutinos.Conclusions} gives a summary and conclusion.

\section{Resonances and Chaotic Diffusion \label{plutinos.Resonances}}

The dynamics of resonant KBOs are characterized by the libration amplitude of their resonant argument, $\phi$.  For the 3:2 and the 2:1 exterior mean motion resonances of Neptune, we will be concerned with the particular resonant arguments,
\begin{eqnarray}
\phi_{3:2} &=& 3\lambda - 2\lambda_N-\varpi, \cr
\phi_{2:1} &=& 2\lambda - \lambda_N-\varpi, 
\label{eq:resargs}
\end{eqnarray}
where $\lambda_N$ is the mean longitude of Neptune, and $\lambda$ and $\varpi$ are the mean longitude and longitude of perihelion, respectively, of a KBO.  

There is one important difference between the 3:2 and the 2:1 resonance librations.  The resonant arguments of particles in most exterior resonances librate about $180^{\circ}$; however, the libration center of a 1:$n$ resonant argument is eccentricity-dependent and bifurcated into multiple asymmetric libration centers \citep{Beauge94,Malhotra96,WM97,PS04}.  For the 2:1 resonance in particular, the asymmetric libration centers appear as a bifurcation of the exact resonant circular periodic orbit at $\phi=180^\circ$; for $e\approx0.1$, the libration centers are near $\pm90^\circ$, and asymptotically approach values near $\phi=\pm 60^{\circ}$ for increasing values of eccentricity~\citep[see figure in][]{Malhotra99}.  A particle whose resonant argument $\phi$ librates about the positive value ($\phi_c$) reaches its perihelion near a longitude that is $\phi_c$ \textit{ahead} of Neptune's position at that time; thus such particles are referred to as ``leading'' librators.  Conversely, a particle whose $\phi$ librates about the corresponding negative value ($-\phi_c$) reaches its perihelion at a longitude that is $\phi_c$ \textit{behind} Neptune's position, and is thus referred to as a ``trailing'' librator.  ``Symmetric'' librators, whose $\phi$ librates about $180^{\circ}$, can continue to exist even when the asymmetric libration centers are present, but they necessarily have large libration amplitudes because they must effectively transition from the leading lobe to the trailing lobe and back to the leading lobe during each libration cycle \citep{Malhotra96}; these are analogous to ``horseshoe orbits'' in the classical three-body problem.

The amplitude of the resonance libration is an important indicator of the stability of a resonant orbit, because a close encounter with Neptune becomes more likely as the particle's resonant argument strays away from the center of libration.  We measure this libration amplitude, $\Delta \phi$, as half the difference between the minimum and maximum excursions of $\phi$.  In the simplest analytical models for resonance, the libration amplitude is a constant of the motion.  However, in reality, $\Delta \phi$ is not an exact constant because additional weak resonances and near-resonances, associated not only with Neptune but with the other planets (primarily Jupiter, Saturn, and Uranus), interact with the dominant resonant perturbation from Neptune and produce quasi-periodic perturbations as well as chaotic behavior \citep[e.g.][]{Lecar}.  To assess the long term stability of resonant KBOs, we are particularly interested in the chaotic evolution that may lead to escape from resonance.  

There are two defining characteristics of a chaotic dynamical system.  One of these is the local exponential divergence of initially nearby particle orbits, and is quantified by the Lyapunov exponent \citep[e.g., see][]{NR00}; however, the relationship between the Lyapunov exponent and escape time is not a simple one.  
The second characteristic of dynamical chaos is the existence of broad frequency bands in the Fourier spectrum of the time series of a dynamical variable (in contrast with the line spectrum for quasi-periodic regular dynamics).  This property has been used to detect the presence of chaos and to measure the sizes of chaotic zones in the dynamics of the major planets of the solar system~\citep[e.g.][]{Laskar90}.  However, this also does not provide a direct estimate of the escape times that are of interest here.  For the purpose of the present paper, a more practical approach approximates the chaotic behavior as a classical diffusion process as follows.
We can define the approximate constants of the motion -- the proper elements -- over a time interval, $\Delta t$, which is long compared to the quasiperiodic perturbations but small compared to the long time of interest. (In a low order analytical perturbation theory, these proper elements are exact constants of the motion, but their constancy is not guaranteed when higher order perturbations are included.)  The changes of the proper elements over successive time intervals can be attributed to chaos when one measures an increase with time of the {\it dispersion} of the proper elements; this is referred to as ``chaotic diffusion''.  
This approach has been developed by \cite{Laskar94} and was adopted by \cite{Morby97} to study the chaotic evolution of Plutinos.  
We adopt this approach in the present work, as it lends itself to a clear analysis and interpretation of the long term changes in the populations of resonant KBOs.  

\begin{figure}[!t]
\begin{center}
\includegraphics[width=12cm]{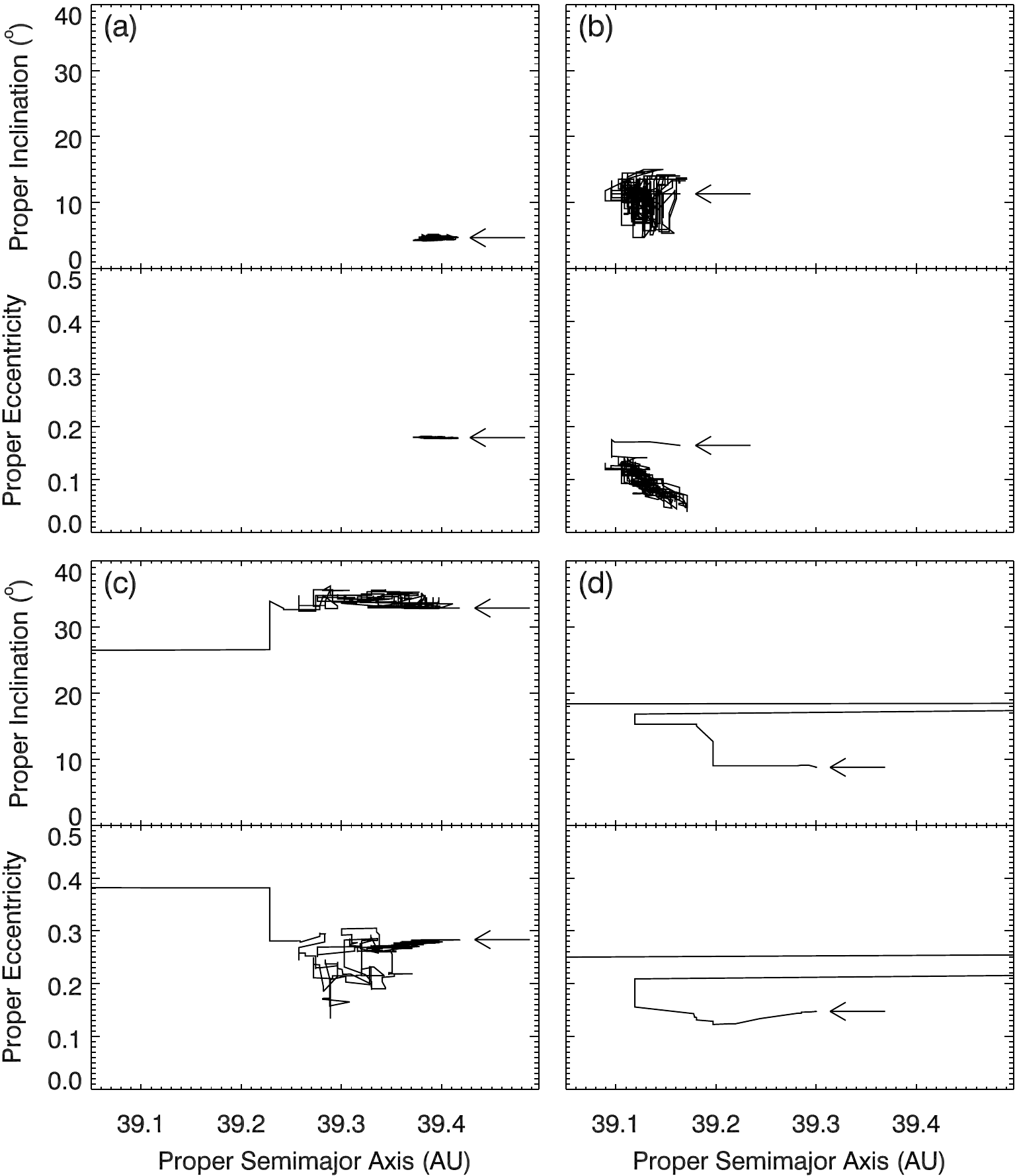}
\figcaption{The evolution of four selected particles from our models illustrates the classes of behavior that we observe: a)~stable resonant orbit with little measurable diffusion, b)~stable resonant orbit which diffuses within resonance, c)~initially resonant orbit which eventually diffuses out of resonance (after 809~Myr in this case), d)~strongly chaotic trajectory which quickly leaves resonance (after 37~Myr in this case).  The minimum proper elements (see Section~\ref{PropElems}) are plotted here, so the right-hand side of each plot is the center of the 3:2~resonance.  Arrows point to the initial values.  \label{DiffusionExamples}}
\end{center}
\end{figure}

Following \cite{Morby97}, we define proper elements that are derived directly from the osculating orbital elements by means of a numerical procedure described in Section~\ref{PropElems}; this procedure removes quasi-periodic variations so that the chaotic behavior can be isolated.  As in \citet{Morby97}'s study of the Plutinos, we observe 4~classes of behavior among the particles in our numerical models, examples of which are shown in \Fig{}~\ref{DiffusionExamples}.  A particle's proper semimajor axis may remain nearly stationary over the integration, either because the particle is on a stable trajectory (an ``invariant torus'' in phase space) or because its chaotic evolution is below the numerical resolution of our model (\Fig{}~\ref{DiffusionExamples}a); secondly, it may evolve over a significant distance in phase space but remain in resonance (\Fig{}~\ref{DiffusionExamples}b); thirdly, it may evolve within the resonance far enough to reach the chaotic zone and thence escape to a non-resonant orbit (\Fig{}~\ref{DiffusionExamples}c); or fourthly, it may be strongly chaotic from the start and escape the resonance almost immediately (\Fig{}~\ref{DiffusionExamples}d).

In order to quantify the chaotic behavior for initial conditions across the resonance phase space, we assume that the chaotic evolution of the proper elements can be approximated as a diffusive process. In a diffusive process, ensembles of particles that start out nearby in phase space drift apart, as their proper elements execute a random walk.  The rate at which they disperse is described by a diffusion coefficient, which we define as $D = \langle(\Delta a)^2\rangle/\Delta t$, where $\Delta a$ is the change in proper semimajor axis over a time interval $\Delta t$, and the mean is taken over the ensemble of particles located in a small volume of phase space.  The diffusion coefficient may vary with location in phase space, in which case particles will tend to spend more time in regions with small diffusion coefficient, and less time in regions with large diffusion coefficient.  
As explained in Section~\ref{PropElems} below, we choose a ``running window'' time interval $\Delta t = 10$~Myr, which is much longer than any resonance libration periods, but much shorter than the $\sim10^3$ Myr timescale of interest for the history of the resonant KBOs.  

\section{Numerical Model and Analysis\label{NumModel}}

\subsection{Pre-runs and Initial Conditions \label{ICs}}

To obtain a sample of initial conditions that comprehensively covers the resonance zone, a series of ``pre-runs'' was performed.  The primary purpose of the pre-runs is to identify values of the initial conditions that result in resonant particles, and then to include only those particles in the full 1-Gyr integration.  This step is necessary because the initial osculating orbital elements do not have a simple relationship with the proper elements that define the parameters of the resonance phase space; this complexity owes to the large-amplitude short-period variations caused by the giant planets (Section~\ref{PropElems}). For the full 1-Gyr runs, we then chose a subset of initial conditions from the pre-runs to ensure adequate coverage of the entire resonance zone, including the deepest zone at very small libration amplitudes; the latter has usually been poorly covered in previous studies.  

The orbital intergrations were performed using the ``Swift-Skeel'' mixed-variable symplectic $N$-body integrator \citep{DLL98,WH91}; we used a step size of 0.5 Earth years.  All runs included the 4 giant planets, whose initial orbital elements (obtained from JPL\fn{http://ssd.jpl.nasa.gov/horizons.html}) corresponded to a starting epoch of 1997 June~1.  The mutual perturbations of the massive planets were fully accounted for.

For the pre-runs, our set of test particles began with 20 eccentricity values, evenly spaced in the interval $e=[.05,.5]$; and 7 inclination values,\fn{In this paper, we measured inclinations with respect to the J2000 ecliptic plane.  The angle between the ecliptic plane and the Laplace plane of the Kuiper Belt (i.e.,~the plane about which orbital planes actually precess) is 1.86$^\circ$ \citep{BP04}, and the mean correction is $<1^\circ$ \citep{TB02}, much smaller than the inclinations discussed in this paper.} evenly spaced in the interval $i=[5^\circ,35^\circ]$.  For Plutinos, an 8th inclination value of $17.5^{\circ}$ was added to increase the resolution in Pluto's vicinity.  Each $\{e,i\}$ bin contained 250 particles, with initial values of the angular elements ($\omega$, $\Omega$, and $M$) randomly distributed in the interval $M=[0^{\circ},360^{\circ})$.  The total number of particles was 35,000 Twotinos, and 40,000 Plutinos.  All particles began with semimajor axis $a=39.35$~AU for Plutinos, $a=47.87$~AU for Twotinos.  The orbits of these particles were then integrated, along with the giant planets, for 150,000~yr; this is several times the typical resonance libration period, and long enough for stable libration of the resonant argument to manifest itself.  At each 2,000-yr interval, we recorded for each planet and test particle the minimum and maximum values of $a$, $e$, and $i$ over that interval, as well as instantaneous values of $\Omega$, $\omega$, and $M$.

For each particle in the pre-run, we track the evolution of the resonant argument, $\phi_{3:2}$ for Plutinos and $\phi_{2:1}$ for Twotinos (see \Eqn{}~\ref{eq:resargs}); and we verify persistence of libration by visual examination of the time series plots of these resonant arguments.  For librating particles, we calculate the libration amplitude of the resonant argument, $\Delta \phi=(\phi_{max}-\phi_{min})/2$. We found that the 75 instantaneous points over the 150-kyr interval were sufficient to characterize the libration.  For Twotinos, we additionally categorized each particle, by visual inspection of the time series of $\phi$, as having stable libration in one of three modes:  leading ($\phi < 180^{\circ}$ at all times), trailing ($\phi > 180^{\circ}$ at all times), or symmetric ($\phi$ librates about $180^{\circ}$).  Particles that did not exclusively exhibit one of these behaviors during the pre-run were discarded as they are of no interest for the long term history of the resonant KBOs.  

Initial particles for the full 1-Gyr integration were then chosen from among the particles that exhibited stable libration in the pre-run.  One in 16 particles with $\Delta \phi > 30^\circ$ was randomly selected.  For particles with $\Delta \phi < 30^\circ$, the probability of selection was increased to one in 4, to enhance the resolution of our results for tightly-bound librators; these particles were weighted by 0.25 in the subsequent analysis.  The result was a set of 1,331 Plutinos and 1,445 Twotinos in the full 1-Gyr integration.  The selected Twotinos are broken down as 479~leading, 636~trailing, and 330~symmetric librators; the weighted totals are 301~leading, 354~trailing, and 330~symmetric.  
Their initial $e$, $i$, and $\phi$ can be seen in \Fig{s}~\ref{P0ResTimes}, \ref{P1ResTimes}, and~\ref{T0ResTimes}.  Overall, librating Twotinos in the pre-run were 31.1\% leading, 33.9\% trailing, and 35.1\% symmetric; Poisson statistics gives a precision of 0.8\% for these proportions. 
There is a hint in Fig{}~\ref{T0ResTimes} that the phase space volume at low values of $e\lesssim0.1$ increases with inclination, indicating a change in the dynamical structure of the 2:1 resonance with inclination; it would be interesting to examine this in a future study with larger numbers of particles.

A final piece of information obtained from the pre-runs is the amplitude of the short-period oscillations in $a,e$, and $i$ caused by the three inner giant planets.  These oscillations occur on timescales as small as tens of years, and are easily averaged over by the 2,000-yr output intervals of the pre-runs.  Therefore, we define for each particle over each output interval a quantity $da=(a_{max}-a_{min})/2$ (and similarly $de$ and $di$). We find that $da, de$ and $di$ remain remarkably constant over the 150-kyr pre-run integration length (see \Fig{}~\ref{Prerun}).  We calculated the mean values of $da$, $de$, and $di$ over the 150-kyr interval for each particle, and adopt these as the amplitudes of the short period perturbations (see Section~\ref{PropElems}).

\begin{figure}[!t]
\begin{center}
\includegraphics[width=15cm]{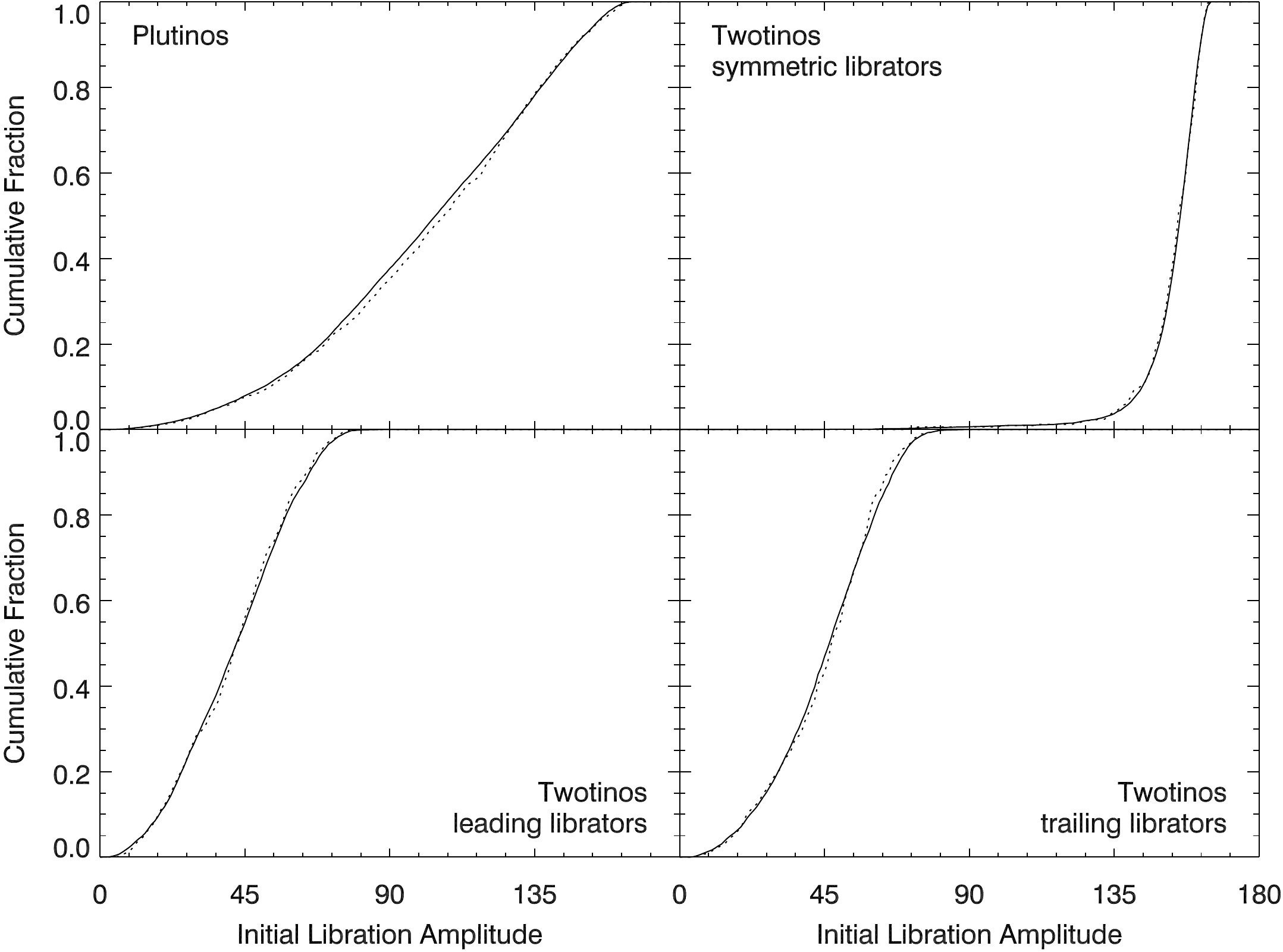}
\figcaption{The solid lines show the distribution of libration amplitudes in the preruns, an unbiased population which reflects relative volumes in resonance phase space.  The dotted lines show the weighted distribution of libration amplitudes in the set of particles selected for our full 1-Gyr integrations.  \label{Weights}}
\end{center}
\end{figure}

The solid lines in \Fig{}~\ref{Weights} show the distribution of resonance libration amplitude, $\Delta \phi$, from our pre-runs.  These distributions obtain from a large number of particles sampling uniformly a wide range of $e$ and $i$, and thus reflect the relative volume of resonance phase space characterized by any given value of $\Delta \phi$---thus we may call it the volumetric distribution of $\Delta \phi$.  Note that this distribution includes all particles that are stable for 150,000~yr (a few libration periods), and thus does not reflect longer-term stability.  From the shapes of these distribution functions, we can infer the relative volumes of phase space at various libration amplitudes.  For Plutinos, we see that relative volumes within the resonance zone are nearly uniformly distributed in the interval $45^\circ \lesssim \Delta \phi \lesssim 160^\circ$, as indicated by the nearly constant slope of the cumulative fraction, while values of $\Delta \phi\lesssim45^\circ$ represent less than $\sim10\%$ of the resonance phase space.  Leading and trailing Twotinos have most of their phase space volume in the range $20^\circ \lesssim \Delta \phi \lesssim 75^\circ$; smaller and large libration amplitudes represent very little resonance phase space volume. For symmetrically-librating Twotinos, the phase space volume is mainly in the range $135^\circ \lesssim \Delta \phi \lesssim 165^\circ$, though amplitudes as low as 90$^\circ$ are possible.

With the initial conditions generated as described above, three numerical integrations were performed for a duration of 1~Gyr each. Run P0 for Plutinos included only 3:2 resonant test particles in the gravitational field of the Sun and the 4 giant planets; similarly, the one Twotino run, T0, included only 2:1 resonant test particles.  A second Plutino run, P1, included Pluto as an additional massive perturber for the 3:2 resonant test particles.  

\subsection{Proper Elements \label{PropElems}}

The osculating orbital parameters of a resonant KBO can vary in a number of different ways, but the dominant variation is due to resonant perturbations of Neptune.  The libration of the resonant argument $\phi$ is accompanied by correlated librations of $a$ and $e$.  In the approximation of the circular restricted 3-body model for resonant orbits (which pertains well to the qualitative characteristics of our particle orbits), the resonant librations of $a$ and $e$ are constrained by the conservation of the action $N$, 
also known as the ``adiabatic invariant''~\citep[see, e.g.,][ch.~8.12]{MD99}, 
\begin{equation}
N = \sqrt{a} \left[ p - (p+1) \sqrt{1-e^2}\cos i \right],
\end{equation}
where $p=1$ for 2:1 resonant particles and $p=2$ for 3:2 resonant particles.  The librations of $a$ and $e$ are $90^\circ$ out of phase with the librations of $\phi$.  
$N$ is conserved under resonant motions only; perturbations from massive bodies other than Neptune, and indeed non-resonant terms in the Disturbing Function due to Neptune, do not conserve $N$.  The resonant variations of these dynamical parameters have characteristic amplitudes and periods which depend on the particle's eccentricity and libration amplitude; the periods are in the range of $10^4$ to $10^5$ years.  Any of these libration amplitudes (e.g., $\Delta \phi$ or $\Delta a$) is diagnostic of how stable the particle's orbit is, with the smallest libration amplitudes usually being the most stable.  However, numerical analysis of the resonant librations is not a simple matter, as other modes of variation occur simultaneously in each particle's evolution, including fast quasi-periodic variations as well as secular perturbations.  

Non-resonant perturbations from the 3~inner giant planets (Jupiter, Saturn, and Uranus) give rise to a suite of fast variations in a resonant particle's orbital elements.  These variations can have significant amplitudes, but their periods are on the order of the period of conjunctions between the disturbing planet and the test particle.  Non-resonant perturbations also arise from Neptune. For the Plutinos and Twotinos, these short period variations have periods ranging from $\sim 12$~yr (owing to Jupiter) to $\sim T_{Nep}/[1-p/(p+1)]$ \citep{MyThesis}; the latter are owed to conjuctions with Neptune, and have periods
493~yr for the 3:2 resonance ($p=2$) or 329~yr for the 2:1 resonance ($p=1$).  The amplitudes of these short-period variations remain very nearly constant and they average to zero over the resonance libration periods.

\begin{figure}[!t]
\begin{center}
\includegraphics[width=7.9cm]{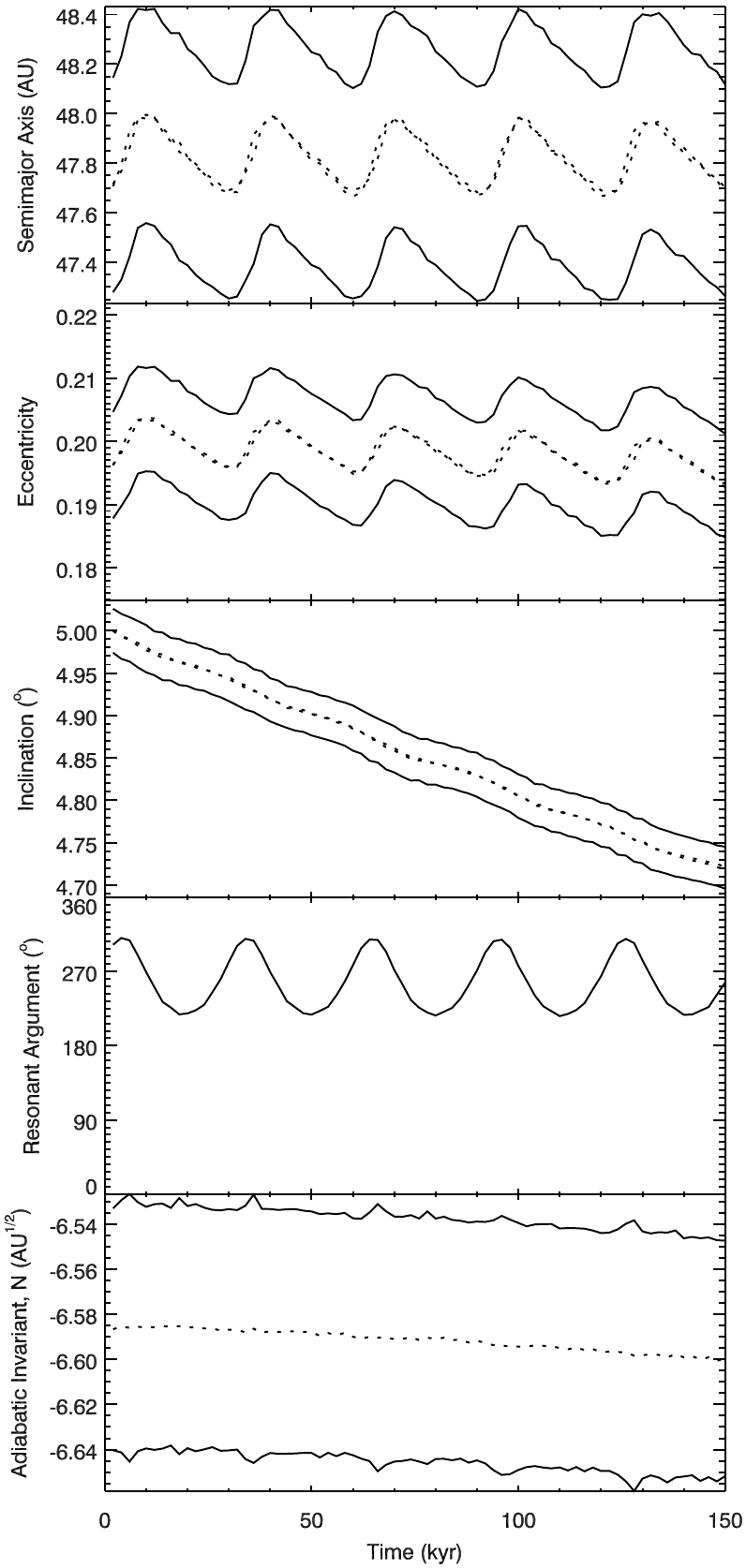}
\figcaption{Pre-run output for one test particle, illustrating the different kinds of variation of the orbital elements as discussed in Section~\ref{PropElems}.  Solid lines are recorded maximum and minimum values of $a$, $e$, and $i$ over 2,000-yr intervals, and instantaneous values of the resonant argument.  Dotted lines {in the top three panels} show the maxima and minima shifted by the constant values $da$, $de$, and $di$, and simply the midpoint value of $N$ in the bottom panel.  Note that this particle is a Twotino, librating in the trailing mode.  \label{Prerun}}
\end{center}
\end{figure}

The giant planets also give rise to long period secular variations in test particle orbits \citep{Knezevic}.  The semimajor axis is not affected by secular variations, but all other orbital elements are.  In the outer solar system, these secular variations occur on $\sim$Myr timescales, much slower than the mean-motion resonant librations.  Thus the latter will continue to move along lines of constant ``adiabatic invariant'' $N$, even as the secular variations cause $N$ itself to oscillate.  

\Fig{}~\ref{Prerun} shows a sample particle from the pre-run (see Section~\ref{ICs}) that demonstrates all of the behaviors described here.  The resonant libration, with a period of 30,000~yr for this particle, is clearly visible in the time evolution of $a$, $e$, and $\phi$, while the evolution of $i$ is dominated by the long-period secular variation.  The secular variation is also perceptible as a gentle downward trend in the eccentricity evolution.  The fast variations are indicated by the difference between the upper and lower solid lines (note that the output timestep for the pre-run is 2,000~yr, small enough that only the fast variations are smoothed over), and the constancy of their amplitude is easily seen by the consonance of the two dotted lines in each panel.  

To tease out the diffusive evolution of particles due to long-term chaos, we adopt the method of \citet{Morby97}, as follows.  In the full 1-Gyr integration, we record for each particle, at 1-Myr intervals, the minimum and maximum values of $a$, $e$, and $i$, over a sliding window with a length of 10~Myr (thus the intervals are [0,10]~Myr, [1,11]~Myr, [2,12]~Myr, etc.).  Additionally, instantaneous values of $\Omega$, $\omega$, and $M$ are also recorded every 1~Myr, so that we can follow directly the evolution of the resonance angles.  Note that the results of this procedure will not explicitly display any of the behaviors described above and seen in \Fig{}~\ref{Prerun}, since all have periods smaller than 10~Myr; rather, these processes are smoothed over, and their amplitudes incorporated into the differences between the recorded maximum and minimum values.  We then remove the amplitude from the fast oscillations induced by the giant planets, by subtracting from the maximum values (and adding to the minimum values) the short-period amplitudes $da$, $de$, and $di$ that were obtained from the pre-runs (see Section~\ref{ICs}).  Finally, we remove secular oscillations from the proper eccentricities by readjusting them to correspond to the midpoint value of the adiabatic invariant\fn{This ``$N$ algorithm'' is a 2-dimensional calculation performed in $a-e$ space, which provides a good approximation of the secular variations in $e$.  The complexity of adding a third dimension to this process precludes a similar calculation of the secular variations in $i$.  However, we note that the latter are generally only a few degrees in amplitude.} $N$ over each 10-Myr interval.  A more detailed  discussion of this so-called ``$N$ algorithm'' is given in~\citet{Morby97}.  

The final proper elements defined by this process are the minimum (which we now refer to as $a_1$, $e_1$, and $i_1$) and the maximum ($a_2$, $e_2$, and $i_2$) values that represent variation due only to the resonant librations. These would be stationary in the long term but for chaotic diffusion.  The evolution of the the proper element set $(a_1,e_1,i_1)$ for four example particles is shown in \Fig{}~\ref{DiffusionExamples}; the chaotic evolution of these particles ranges from very stable (little change in the proper elements) to strongly chaotic (large and rapid changes in the proper elements). To quantify the chaotic diffusion, we determine the variation $\Delta a_1$ between each pair of consecutive sliding-average time windows described in the previous paragraph.  Then we calculate at each location in phase space the chaotic diffusion coefficient $D = \langle(\Delta a_1)^2\rangle/\Delta t$ (see Section~\ref{plutinos.Resonances}), where $\Delta t=10$ Myr, with the average taken over all particles at all times at that phase-space location.  We note that this calculation could equivalently have been done with $a_2$.  With this procedure, we find values of $D$ in the range $10^{-5}$--$10^{-2}$ AU$^2$/Myr for the Plutino and Twotino particles in our 1 Gyr simulations.

\begin{table}[!t]
\begin{center}
\caption[Rms deviations of proper planetary orbital elements]{Rms deviations of proper orbital elements for the giant planets over all 1-Gyr integrations described in this paper \label{RmsDev}}
\vspace{0.3in}
\begin{tabular}{ l r @{.} l r @{.} l r @{.} l }
\hline 
\hline
& \multicolumn{2}{c}{$\sigma_a$ (AU)} & \multicolumn{2}{c}{$\sigma_e$} & \multicolumn{2}{c}{$\sigma_i$} \\
\hline
Jupiter   & 0&00003 & 0&00004 & 0&$001^\circ$ \\
Saturn    & 0&00015 & 0&00020 & 0&$004^\circ$ \\
Uranus    & 0&0014 & 0&00054 & 0&$018^\circ$ \\
Neptune   & 0&0040 & 0&00032 & 0&$006^\circ$ \\
\hline
\end{tabular}
\end{center}
\end{table}

To calibrate the chaotic diffusion rates of the particles, we can compare them with the chaotic diffusion rates for the planets, which we expect to be relatively small.  Numerical experiments have shown the evolution of the giant planets to be regular or only very weakly chaotic \citep[and references therein]{Hayes08}, and their orbital elements remain within narrow bounds over multi-gigayear timescales \citep{Laskar97a,IT02} so that their proper elements ought to be very close to stationary. Table~\ref{RmsDev} shows the rms deviations of the proper elements for each of the giant planets over all of our 1-Gyr integrations. (Our values for the first three planets are in agreement with those quoted by \citet{Morby97}, while our values for Neptune are a factor of 3 to 5 larger. We do not know the source of the latter discrepancy, but the detailed study by \citet{Hayes08} suggests that such differences are not surprising in practically similar long-term integrations of the outer solar system that may differ by adopting planets' initial conditions at different epochs and/or use different orbit integrators.) The magnitudes of the rms variations of the planets' elements provide a measure of the numerical resolution of our study for the detection of chaotic diffusion. The largest rms deviation, $\sigma_a$ for Neptune, is equivalent to a diffusion coefficient $D \approx 10^{-8}$~AU$^2$/Myr, which is three orders of magnitude smaller than the smallest diffusion coefficient we find for the resonant particles. 

\section{Results \label{plutinos.Results}}
Initial parameters and resonance residence times for particles in the three full integrations are shown in \Fig{s}~\ref{P0ResTimes}---\ref{T0ResTimes}.  We define a Plutino as having escaped the resonance as soon as $a_1$ falls below 38.77~AU or $a_2$ exceeds 40.17~AU; for Twotinos, the corresponding limits are 47.16~AU and 48.56~AU.  Both of these intervals are $\pm 0.7$~AU from the central resonant value; they are over-generous in that all resonant particles remain within the intervals, while some particles may remain briefly within the intervals despite being no longer in resonance.  We find that once a particle has left the resonance for the first time, it spends very little time (on average, a few percent of each particle's dynamical lifetime) back in the resonant range of semimajor axis.  Therefore, we consider a particle to be permanently non-resonant once it has first left the resonant range.

Maps of the dynamical diffusion rate within the resonant region for each integration are shown in \Fig{}~\ref{DiffMaps}.  For these maps, we divided the proper element planes, $(a,e)$ and $(e,i)$, into small bins, and calculated the average value $\bar D$ of the diffusion coefficient at each phase-space location, averaging over all particles at all times whose initial proper elements lie in that bin at the beginning of a sliding-average time window (see Section~\ref{NumModel}). The grey scale maps show how these average diffusion coefficient values vary across the resonance phase space.  We have also defined the ``characteristic diffusion time" associated with the chaotic diffusion, $\tau=L^2/\bar D$, where we use $L=0.2$~AU as a characteristic lengthscale (half-width) for the resonant region; the scale bar in \Fig{}~\ref{DiffMaps} shows the grey scaling for both $\bar D$ and $\tau$.

Finally, in \Fig{}~\ref{FracLeftPlot}, we show plots of the fraction of particles that survive in resonance as a function of time, and the distribution of the characteristic diffusion time.

These results are discussed in detail below for the Plutinos and Twotinos separately.

\begin{landscape}

\begin{figure}[!t]
\begin{center}
\includegraphics[width=21.5cm]{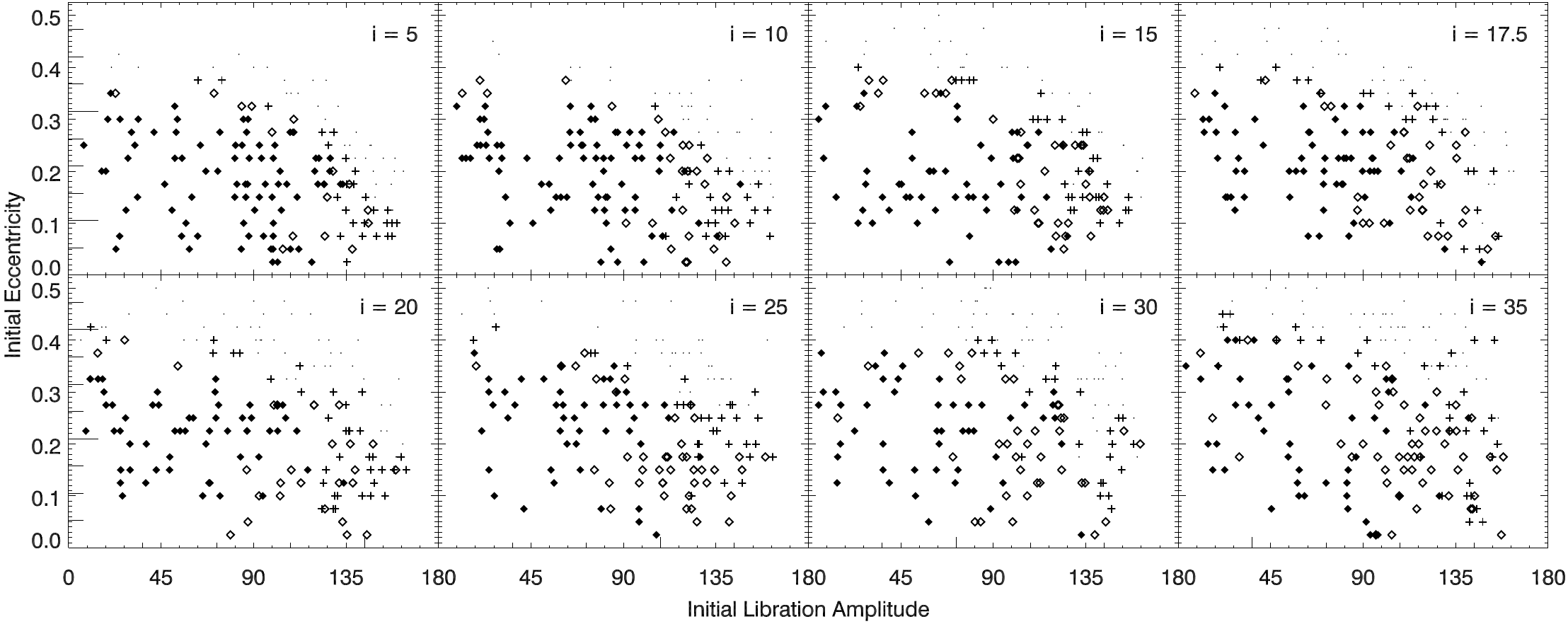}
\figcaption{Initial parameters and outcomes for test particles in run P0.  Solid diamonds indicate particles that remained in resonance for the entire 1~Gyr integration, open diamonds indicate particles that left resonance after 100~Myr, crosses indicate particles that left resonance between 10~Myr and 100~Myr, dots indicate particles that left resonance in less than 10~Myr.  \label{P0ResTimes}}
\end{center}
\end{figure}

\begin{figure}[!t]
\begin{center}
\includegraphics[width=21.5cm]{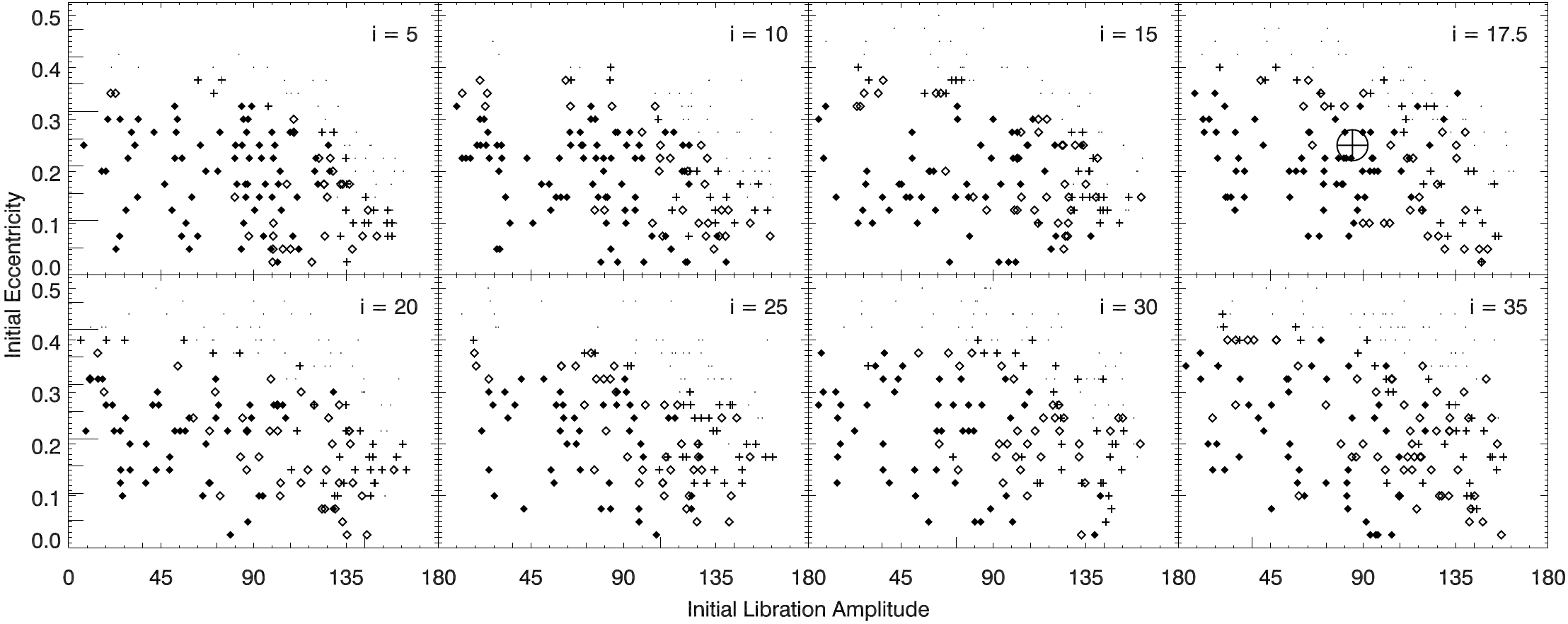}
\figcaption{Initial parameters and outcomes for test particles in run P1.  Symbols as in \Fig{}~\ref{P0ResTimes}.  The large circle with inscribed cross represents Pluto. \label{P1ResTimes}}
\end{center}
\end{figure}

\begin{figure}[!t]
\begin{center}
\includegraphics[width=21.5cm]{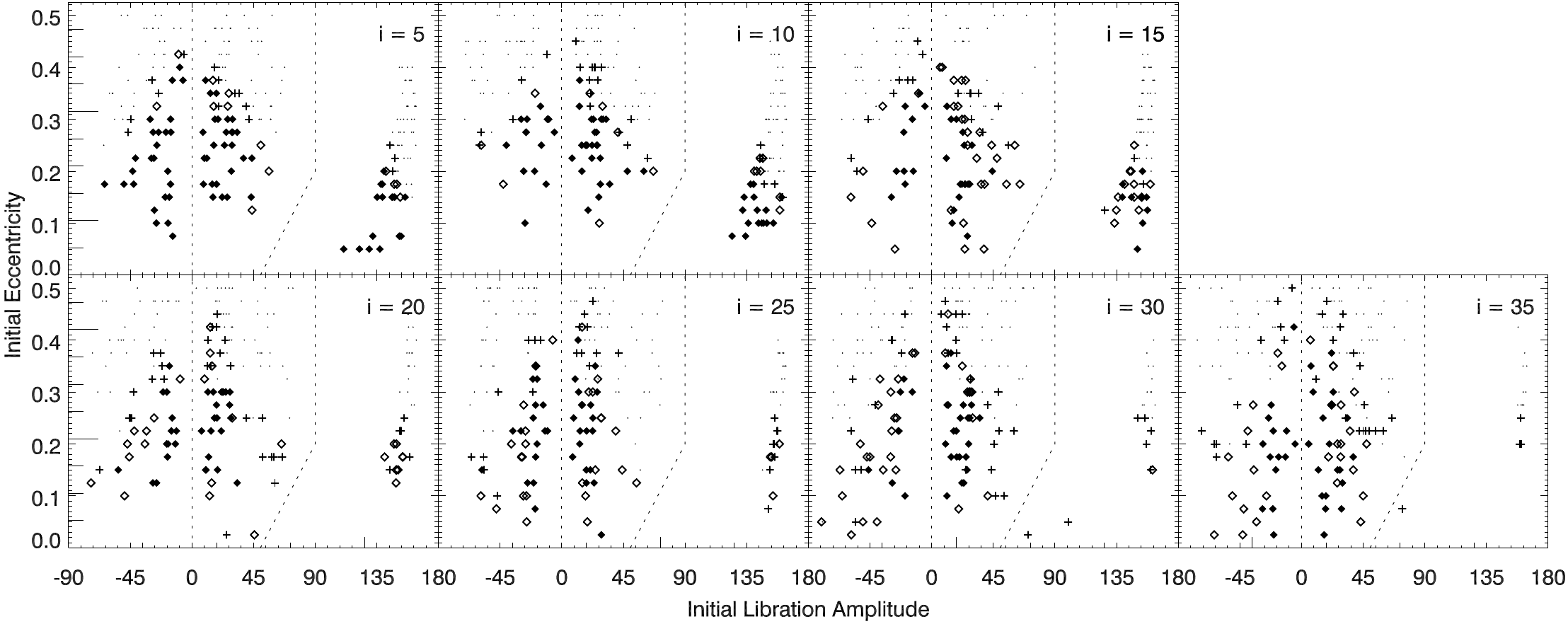}
\figcaption{Initial parameters and outcomes for test particles in run T0.  Dotted lines divide the diagram into three panels, leading librators on the left-hand side, trailing librators in the center, and symmetric librators on the right-hand side.  Symbols as in \Fig{}~\ref{P0ResTimes}.  \label{T0ResTimes}}
\end{center}
\end{figure}

\end{landscape}

\subsection{Plutinos \label{PlutinoResults}}

\begin{figure}[!t]
\begin{center}
\includegraphics[width=16cm]{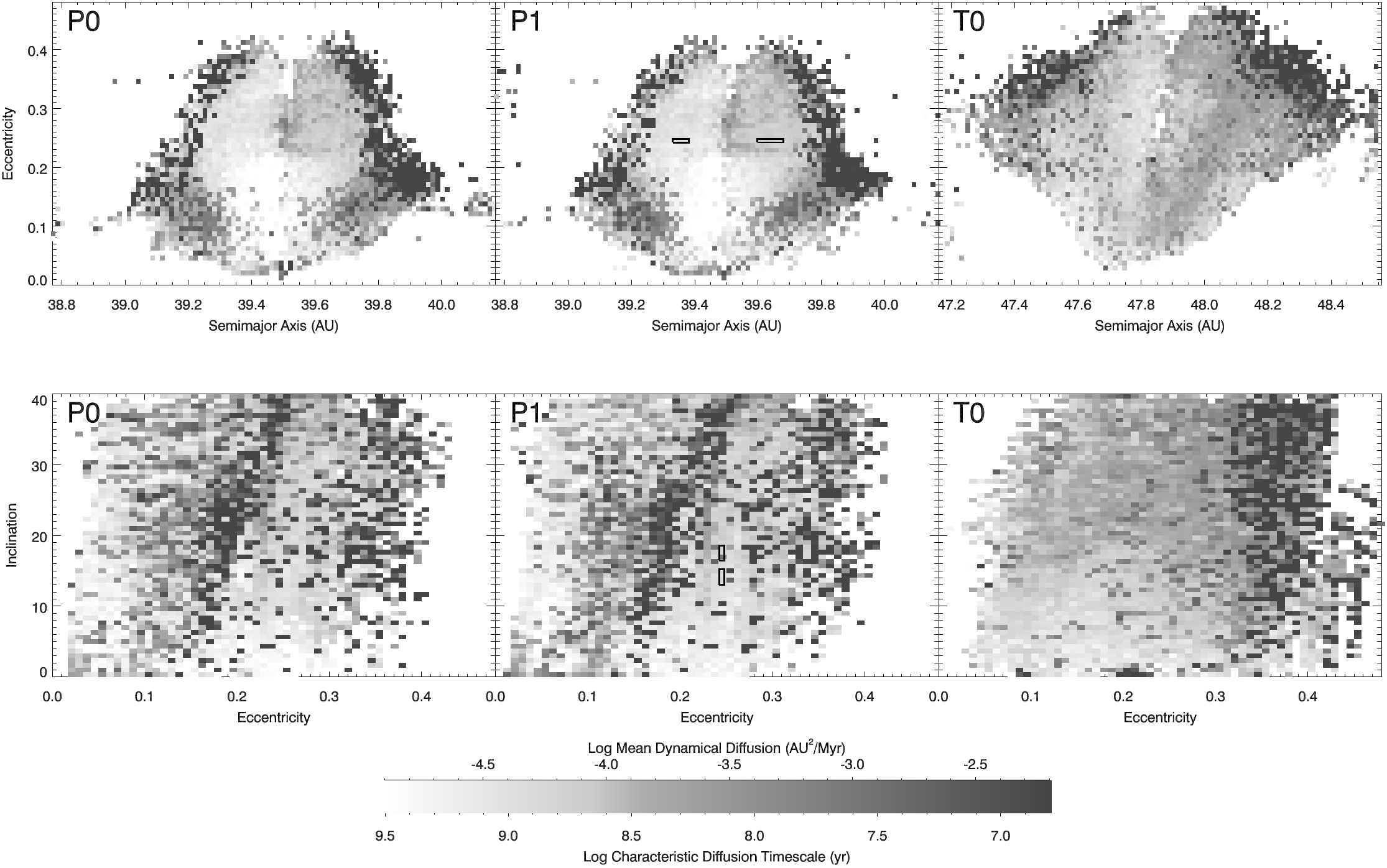}
\figcaption{Maps of mean dynamical diffusion coefficient $\bar{D}$. Also noted on the scale bar is the characteristic time for diffusion associated with each diffusion coefficient, $\tau=L^2/\bar{D}$, where we use $L=0.2$~AU as a characteristic lengthscale (half-width) for the resonant region. The two small boxes in the P1 plots show the extent of the variation in Pluto's minimum and maximum proper elements, respectively. \label{DiffMaps}}
\end{center}
\end{figure}

\begin{figure}[!t]
\begin{center}
\includegraphics[width=12cm]{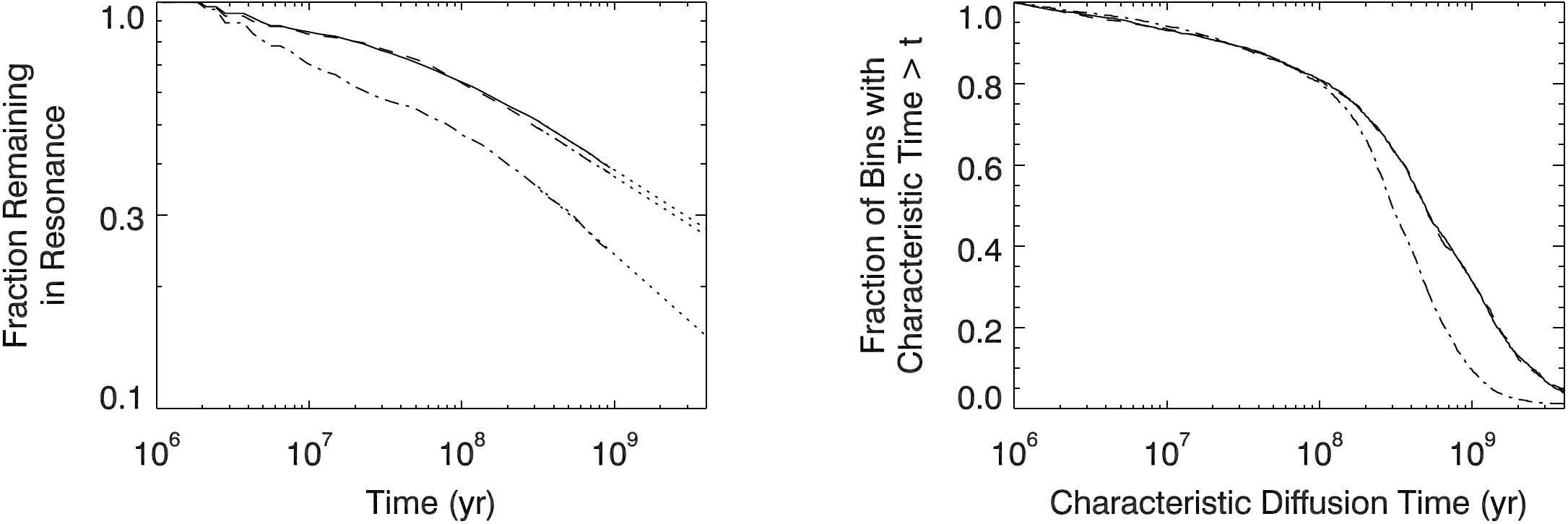}
\figcaption{(a)~Fraction of particles remaining in resonance, as a function of time, for runs P0 (solid), P1 (dashed), and T0 (dot-dash). Dotted lines show power-law fit to the last 0.5 Gyr of data, with extrapolation to 4~Gyr. (b)~Relative frequency of characteristic diffusion times from maps in \Fig{}~\ref{DiffMaps}. \label{FracLeftPlot}}
\end{center}
\end{figure}

In \Fig{s}~\ref{P0ResTimes} and~\ref{P1ResTimes}, we see that long-term stable resonant orbits for Plutinos are prevalent at small $\Delta \phi$ and $e$ values, and only slightly less common at higher inclinations.  In the $a$--$e$ planes (upper panels in \Fig{}~\ref{DiffMaps}), the lowest diffusion values are at $a \sim a_{res}$ and low $e$.  In the $e$--$i$ planes (lower panels in \Fig{}~\ref{DiffMaps}), low diffusion rate is found at low $e$ for the entire range of inclinations, and the Kozai resonance --- in which the argument of pericenter librates about 90$^\circ$, thus enhancing stability \citep{Kozai} --- is visible as a band of lower diffusion rate just to the right of the highest diffusion rates ($0.2 < e < 0.3$).

By comparing the P1 results with those for P0, the effects of Pluto can be assessed.  Comparing \Fig{s}~\ref{P0ResTimes} and~\ref{P1ResTimes}, a number of particles become less stable when Pluto is present, but many also become more stable; Pluto's presence brings only a 3\% net decrease in the mean particle lifetime of a particle in the resonance.  Furthermore, very little difference can be discerned between the diffusion maps for the two runs (\Fig{}~\ref{DiffMaps}).  However, one noticeable effect of Pluto is that the number of particles remaining in the Kozai resonance (identified by tracking the libration of $\omega$) for the entire 1~Gyr is 13\% lower for run P1 as compared with run P0.  Because Pluto also occupies the Kozai resonance, Kozai librators will tend to have orbital elements similar to Pluto's.  Thus, their encounters with Pluto produce greater perturbation because of the lower encounter velocities \citep{Opik}.  There is also a significant number of particles (10\% of the total) that experience ``Trojan'' behavior with respect to Pluto at one time or another (i.e., either libration or slow circulation of $\lambda-\lambda_P$), though few particles (0.5\% of the total) experience long-term Trojan libration.  

Pluto's overall effect on our test particle Plutinos is most clearly seen in the rate at which particles leave the resonance (\Fig{}~\ref{FracLeftPlot}a).  At the end of the 1-Gyr integration, 39\% of the resonant particles in P0 remain in resonance, compared with 37\% in P1.  The percentage projected to remain after 4~Gyr is also only slightly less for P1 (27\%) than for P0 (28\%), and the loss function exponents (Section~\ref{FracLeft}) are similar ($b_{P1}=-0.554$, while $b_{P0}=-0.556$).

\subsection{Twotinos \label{TwotinoResults}}

In \Fig{}~\ref{T0ResTimes} we see that long-term stability in run T0 is correlated with smaller $\Delta \phi$ values; but here we find (in some contrast with the Plutinos) that moderate eccentricities ($0.1 < e < 0.3$) are more likely to be stable than low eccentricities, and that stability falls off visibly with increasing inclination.  For inclinations greater than about $15^\circ$, we find no long term stable symmetric librators. These patterns are also seen in the diffusion maps (\Fig{}~\ref{DiffMaps}), where we also note that the region of stability spreads over a somewhat wider interval in semimajor axis than is the case for Plutinos.  An increase in stability associated with the Kozai resonance is discernible in the $e$--$i$ plane (lower T0 panel in \Fig{}~\ref{DiffMaps}) as a faint band near $e \sim 0.4$, but it appears to be much less prominent in Twotinos than in Plutinos.  

As shown in \Fig{}~\ref{FracLeftPlot}a, 24\% of particles in run T0 remain in resonance after 1~Gyr, about two-thirds the fraction found above for Plutinos, and the loss function exponent (Section~\ref{FracLeft}) is steeper ($b_{T0}=-0.768$), yielding only a projected 15\% remaining after 4~Gyr.  This indicates that the 2:1 resonance has less overall long-term stability than the 3:2.

\subsection{Resonant Population Decay \label{FracLeft}}

\Fig{}~\ref{FracLeftPlot}a shows, as a function of time, the fraction of particles in each run that have not yet escaped the resonance.  For all runs, a downturn or ``elbow'' can be seen at about 100~Myr.  This was also noted by \citet{Morby97} in his study (which corresponds to our run P0), though the effect is more modest in our results.  This ``elbow'' can be explained if the volume of phase space with characteristic diffusion times in the range $10^8 < \tau < 10^9$~yr were much larger than the volume with shorter diffusion times, so that particles would begin escaping from the former region only after $10^8$~yr.  This explanation is supported by the plot in \Fig{}~\ref{FracLeftPlot}b, which shows the distribution of dynamical diffusion times: we see that $\sim80\%$ of the resonance phase space volume has characteristic diffusion time in excess of $10^8$ yr and shorted diffusion times are found in only 20\% of the resonance phase space.

We extrapolate the fraction, $f_{res}$, of remaining resonant particles using a functional form $f_{res} \propto t^b$, which is preferred by \citet{Morby97} on theoretical grounds, although other functional forms would also fit our data, such as  $f_{res} \propto \log t$ \citep{HW93}.  For each run, we have made a least-squares fit to the last 0.5 Gyr of our results to find the loss function exponent $b$, which we use to predict the fraction of particles remaining after 4~Gyr.  For Plutinos, our best-fit exponent is $b_{P0} = b_{P1} = -0.55$; this is similar to \citeauthor{Morby97}'s estimate of $b = -0.5$. For Twotinos (which \citeauthor{Morby97} did not investigate), our best-fit exponent is steeper, $b_{T0}=-0.77$.

The uncertainty in the loss function exponent $b$, and in the extrapolated fraction of particles remaining after 4~Gyr, can be estimated using standard error-propagation techniques.  The standard deviation of the fitted data from the model (a linear fit in log-log space) is no more than 2\% of the range covered by the data.  The uncertainties for our quoted values of $b$ are in the third significant figure.  The uncertainties (calculated from the covariance matrix, and thus assuming that our choice of a linear model is correct) for our quoted fractions of particles remaining after 4~Gyr, extrapolated from our 1-Gyr simulations, are in the fourth significant figure, though we choose to quote only two significant figures. 

\subsection{Resonance Escapees \label{AfterEscape}}

Particles that escape from resonance in our simulations enter other dynamical populations.   The escaped particles spend roughly equal amounts of time as Centaurs and as scattered disk objects (SDOs).  (The definitions of ``Centaur'' are somewhat variable in the literature; we adopted the definition of \citet{Chiang07} for these populations: objects that are neither Resonant nor Classical KBOs are Centaurs if their perihelion is interior to Neptune's orbit, and SDOs otherwise.)  Escaped Plutinos spend somewhat more time (52\% in run P1, 55\% in run P0) as Centaurs, while escaped Twotinos spend somewhat more time (52\%) as SDOs.  Combined, escaped Plutinos and Twotinos spend less than 1\% of their time as classical Kuiper Belt objects (CKBOs)---that is, with mean eccentricity $\langle e \rangle<0.2$ and mean Tisserand parameter with respect to Neptune $\langle T_N \rangle>3$.  We find that  $\sim 40\%$ of resonance escapees survive to the end of our 1-Gyr integration.  These typically survive as SDOs, their stability often enhanced by ``resonance sticking"; this may involve a long-term stay in a single resonance, or several resonances may be visited \citep{LD97}.  

If a particle's perihelion migrates inside the orbit of Neptune while it is unprotected by a resonance, it then enters the population of Centaurs \citep{TM03,HEB04a,diSisto07}.  Centaur dynamics are dominated by scattering due to the giant planets.  Centaur orbital elements tend to diffuse nearly evenly throughout the planet-crossing region of parameter space.  We find that  27\% 
of particles that do not survive the 1-Gyr integration enter the inner solar system (that is, to heliocentric distance $r<5$~AU); this is consistent with previous results (e.g.,~\cite{TM03}).  We also see in our simulations that resonant KBOs make a significant contribution to the SDO population, as well as directly to the Centaurs.

\section{Discussion \label{Discussion}}

\subsection{Comparison with Previous Work}

\citet{Morby97} explored chaotic diffusion in the 3:2 resonance using methods very similar to ours.  We observe all of the dynamical behaviors that he describes, including objects which slowly diffuse from the inner parts of the resonance phase space, escaping only after several billion years.  We also reproduce fairly well his map of chaotic dynamical diffusion in the 3:2 resonance.  We have expanded on his work by adding greater resolution, greater coverage of particles deep in the resonance, and also by exploring the influence of Pluto; we have also extended the study to the 2:1 resonance. We  confirm Morbidelli's observation of an increase in the rate of escape from resonance after $t \sim 100$~Myr (\Fig{}~\ref{FracLeftPlot}a), and note that similar behavior appears in the 2:1 resonance as well.  We ascribe this to a relatively large volume in phase space that has characteristic diffusion times on that order (\Fig{}~\ref{FracLeftPlot}b). 

\citet{YT99} investigated a semi-analytical model based on a simplified resonance Hamiltonian, to explore the dynamics of Pluto and the Plutinos.  All Plutinos in their model experienced significant interactions with Pluto, but their initial conditions were restricted to particles with orbits very similar to Pluto's.  In our more diverse sample, we do observe a set of behavioral classes that is similar to what \citeauthor{YT99} described, including persistent Trojan behavior (``tadpole'' and ``horseshoe'' orbits, and orbits that transition between them), particles with a slow circulation of $\lambda-\lambda_P$, and others that experience only intermittent Pluto Trojan behavior.  About 24\% of particles in our simulation experience at least intermittent commensurability with Pluto, but only 7\% experience persistent circulation of $\lambda-\lambda_P$, and less than 1\% are in tadpole or horseshoe orbits.

\citet{NR00,NR01} investigated the stability of the 3:2 and 2:1 resonances, respectively.  They used the digital filter method of obtaining proper elements, and they measured resonance stability using the Lyapunov exponent and the minimum distance from Neptune, rather than computing the dynamical diffusion coefficients as we have.  Still, our maps of resonance stability have broad agreement in the overall shape of the resonant region of phase space, with the most stable regions of the resonances centered on the resonant values of $a$, and an additional zone of stability associated with the Kozai resonance.  

\citet{NRF00} performed a short-term numerical integration of the known Plutinos in order to determine their present proper orbital elements, including the resonance libration amplitude $\Delta \phi$.  They found a surprising lack of observed Plutinos deep in the 3:2 resonance (i.e. at small $\Delta \phi$), as well as at moderate eccentricities (i.e., near $e_P$) in the Kozai resonance, which they attributed to the effects of Pluto. They also carried out numerical integrations of particles with initial eccentricities and/or inclinations very close to those of Pluto, showing that Pluto ejects a large fraction which would otherwise be stable.  We find a much smaller influence of Pluto upon the Plutinos, likely due primarily to \citeauthor{NRF00}'s reliance on low-amplitude librators---all of their particles begin with $\Delta \phi < 20^\circ$, and most with $\Delta \phi < 10^\circ$---which, as they demonstrate, are much more likely to interact with Pluto.  By contrast, our pre-runs (which do not include Pluto) indicate that less than 2\% of phase space is represented by such low-amplitude librators, a fraction reflected in the initial conditions of our models.  Consequently, we find Pluto's overall effects to be on the order of a few percent increase in objects lost from the resonance over 4~Gyr, rather than 50\%.  This large difference is mainly because we adopt an initial libration amplitude distribution which is proportional to the phase space volume at different libration amplitudes. 

Our present work is a condensed but updated version of the study by \citet{MyThesis}, who additionally investigated the effects of a massive perturber embedded in the 2:1 mean-motion resonance.  The earlier work contained an error in the planetary initial conditions and also used a less robust process for selecting initial conditions for the test particles.  \citet{MyThesis} found a reduced impact of Pluto upon the Plutinos compared to previous work, but here we find Pluto's role even further reduced, likely due primarily to truly random selection of particles from the pre-run to avoid bias towards low-$\Delta \phi$ librators, and secondarily to randomly distributed initial values of $\omega$ and $\Omega$ rather than constraining them to equal Pluto's initial values.  

\subsection{Population Estimates \label{PopEst}}

\citet{Kavelaars08} estimated that Plutinos are currently $\sim 20\%$ as abundant as classical Kuiper Belt objects.  \citeauthor{Kavelaars08} do not give an estimate for the current population of Twotinos, but \citet{CJ02} estimated that Plutinos are $\sim 3$ times as abundant as Twotinos.  Using our results from Sections~\ref{PlutinoResults} and~\ref{TwotinoResults} that 37\% of the 3:2 resonance population survives for 1~Gyr (run P1) and 24\% survives in the 2:1 resonance (run T0), and extrapolating the loss function curves to 4~Gyr (Section~\ref{FracLeft}), we estimate that the Plutino population has decayed $\sim 73\%$ over the past 4 Gyr and the Twotino population has decayed $\sim 85\%$ over that time period. (The uncertainties in these estimates are small -- see \Fig{}~\ref{FracLeftPlot}a and discussion in Section~\ref{FracLeft}; future work could check these estimates by means of full 4-Gyr simulations.)  In contrast with their present abundance, the Plutinos may have formerly been as much as $\sim 75\%$ 
as abundant as CKBOs, and Twotinos as much as $\sim 45\%$ as abundant. 
Thus the resonant population comprised a much larger fraction of the Kuiper Belt 4~Gyr ago, likely more than half of the total (classical + resonant).  
Furthermore, the Plutino/Twotino ratio 4~Gyr ago was about half what it is today.  
These relative population estimates for 4 Gyr ago are consistent with models that predict the two populations to have comparable numbers at the conclusion of Neptune's migration~\citep[e.g.,][]{Malhotra95,HM05}.\fn{It is possible that the early Kuiper Belt history may have included rapid loss from the strongly unstable regions of the resonances, and such loss is unconstrained by the presently observed populations.}

The above estimates and conclusions are based on the assumption of an initially volumetric distribution of $\Delta \phi$, meaning that the frequency of a particular value of $\Delta \phi$ is proportional to the volume in phase space characterized by that value (particles that do not survive the first 1~Myr of the full integration are not counted, as they are not likely to have survived any process by which the Kuiper Belt could have been formed).  It is unclear whether this assumption is robust; current alternative models of Kuiper belt structure formation have not noted the resonance libration amplitude distributions. If, instead of the volumetric distribution, only the very stable particles survive the formation process to become the initial resonant populations, then particle escapes overall would be much less common; consequently, the fraction of resonant particles that survive to the present age of the solar system would be much larger, and neither of the two conclusions above would be very robust.  For illustration, we consider the relatively extreme case that initially only the small amplitude librating orbits, $\Delta\phi<30^\circ$, were populated; in this case, we estimate from our simulations that approximately 67\%, 56\% and 38\% of the populations in our model runs P0, P1 and T0, respectively, would survive to 4 Gyr (instead of approximately 28\%, 27\%, and 15\%, as described above).  In this case, the projected primordial resonant populations and the Plutino/Twotino ratio are significantly different, and, as discussed above, Pluto's perturbing effects are significantly larger. These considerations show that models of Kuiper Belt structure formation should pay attention to the resonance libration amplitude distributions, as the latter are critical in determining the history of the resonant populations; conversely, the resonance libration amplitude distributions may be critically useful in constraining alternative theoretical models.  This is a possible direction for future research.

\section{Conclusions \label{plutinos.Conclusions}}

We carried out extensive numerical integrations to probe the characteristics of the 3:2 (Plutinos) and 2:1 (Twotinos) mean motion resonances with Neptune.  In each case we obtained a representative uniform sample of initial conditions in the resonance phase space, and integrated these for 1~Gyr.  Our results can be summarized as follows:

\begin{itemize}
\item[1.] In both resonances, for orbits that are stable on resonance libration timescales ($\sim 10^5$~yr), the resonance phase space volume is described by the following distributions of the libration amplitudes ($\Delta \phi$).  For Plutinos, $\Delta \phi$ has uniform distribution between $45^{\circ}$ and $160^{\circ}$; only a small fraction, $<5\%$, of the phase space volume has $\Delta\phi<30^\circ$.  For Twotinos, symmetric librators have a nearly uniform distribution of $\Delta\phi$ between $145^{\circ}$ and $165^{\circ}$, while asymmetric librators (both leading and trailing) have a nearly uniform $\Delta\phi$ distribution between $20^{\circ}$ and $75^{\circ}$; smaller and larger libration amplitudes represent $<10\%$.
\item[2.] The presence of Pluto has only modest effects on the Plutino population.  It disrupts the stability of objects in the Kozai resonance and causes some particles to undergo libration in a 1:1 (``Trojan'') resonance.  The overall fraction of Plutinos that survive for 1~Gyr is 39\% when Pluto's perturbations are neglected, and 37\% when they are included.  Projected to 4~Gyr, these two surviving fractions are 28\% and 27\%, respectively (\Fig{}~\ref{FracLeftPlot}a).  If, however, the initial distribution of $\Delta \phi$ were heavily weighted towards low-amplitude librators, Pluto's effects would likely be more important. 
\item[3.] The 2:1 resonance has weaker overall long-term stability than the 3:2; only $\sim 24\%$ of Twotinos survive in our 1~Gyr integration, and only $\sim 15\%$ are projected to survive for 4~Gyr (\Fig{}~\ref{FracLeftPlot}a).
\item[4.] The population decay rates obtained in our models indicate that, 4~Gyr ago, resonant KBOs made up more than half of the Kuiper Belt, and the Plutino/Twotino ratio was closer to unity than it is now.  These estimates depend on our assumption of an initial distribution of particles proportional to the local phase space volume in the resonance and to the extrapolation from our 1-Gyr integrations.  These population ratio estimates for the 4-Gyr old Kuiper belt are roughly consistent with simple models of Neptune's migration and resonance sweeping of the Kuiper Belt \citep{Malhotra95}, and provide a useful constraint for more detailed models \citep{MB04,MLG08}.
\end{itemize}

\acknowledgements{We thank H.~Levison and an unnamed reviewer for help in improving this manuscript.  We thank M.~Hedman for helpful discussions.  We acknowledge research support from NASA Outer Planets Research grants NNG05GH44G and NNX08AQ65G.  MST additionally acknowledges support from NASA Planetary Geology \& Geophysics grant NNX08AL25G.}

\end{document}